\documentclass[twocolumn,preprintnumbers,amsmath,amssymb,prc]{revtex4}
\usepackage[dvips]{graphicx}

\newcommand{\tr}{\mathrm{tr}}
\newcommand{\Pcal}{\mathcal{P}}
\newcommand{\xt}{\boldsymbol{x}_\perp}
\newcommand{\yt}{\boldsymbol{y}_\perp}
\newcommand{\del}{\boldsymbol{\partial}_\perp}
\newcommand{\LQCD}{\Lambda_{\text{QCD}}}
\newcommand{\half}{{\tfrac{1}{2}}}

\begin{document}
\title{Erratum: Initial fields and instability in the classical model
  of the heavy-ion collision}
\author{Kenji Fukushima}
\affiliation{Yukawa Institute for Theoretical Physics,
 Kyoto University, Kyoto 606-8502, Japan}
\begin{abstract}
 We correct a mistake in the analytical expression for the energy
 density given in Phys.\ Rev.\ C {\bf 76}, 021902 (2007)
 [arXiv:0704.3625 [hep-ph]].  The expression should be multiplied by
 $16$.  One question then arises; how could it be possible to explain
 this difference between the analytical and numerical results in the
 same model if both are correct?  We find a subtle problem in the
 treatment of the randomness of the color source along the
 longitudinal direction and the treatment of the longitudinal extent
 of the color source.
\end{abstract}
\maketitle


  The initial energy density given in Ref.~\cite{Fukushima:2007ja}
should be multiplied by $16$, that is, Eq.~(6) should be corrected as
\begin{equation}
 \frac{g^2}{(g^2\mu)^4}\cdot \Bigl\langle 2\tr\bigl(B^\eta_{(0)}
  \bigr)^2\Bigr\rangle = \frac{1}{2}N_c(N_c^2-1)\sigma^2 \,,
\end{equation}
and, accordingly, Eq.~(8) should be corrected as
\begin{equation}
 \frac{g^2}{(g^2\mu)^4}\cdot \varepsilon_{(0)} = 12\sigma^2 \,.
\end{equation}
Also, not only the $\tau^0$-order terms but the $\tau^2$-order terms
missed the same factor $16$.  Equation~(11), therefore, should be
\begin{align}
 &\frac{g^2}{(g^2\mu)^4}\cdot 2\Bigl\langle 2\tr\bigl(B^\eta_{(2)}
  B^\eta_{(0)}\bigr)\Bigr\rangle
 =\frac{g^2}{(g^2\mu)^4}\cdot 2\Bigl\langle 2\tr\bigl(E^\eta_{(2)}
  E^\eta_{(0)}\bigr)\Bigr\rangle \notag\\
 &=-\frac{1}{2}N_c(N_c^2-1)\,\sigma\cdot\chi \,,
\end{align}
and Eq.~(13) should be
\begin{align}
 &\frac{g^2}{(g^2\mu)^4}\cdot \Bigl\langle 2\tr\bigl(B^i_{(2)}
  B^i_{(2)}\bigr)\Bigr\rangle
 =\frac{g^2}{(g^2\mu)^4}\cdot \Bigl\langle 2\tr\bigl(E^i_{(2)}
  E^i_{(2)}\bigr)\Bigr\rangle \notag\\
 &=\frac{1}{4}N_c(N_c^2-1)\,\sigma\cdot\chi \,.
\end{align}
As a result, the estimate for the initial energy density in Eq.~(14)
(which had a typo; $\pi$ in the square brackets should be $\half$),
its resummed form in Eq.~(15), and the compact formula in Eq.~(16) in
the continuum limit should be corrected respectively as
\begin{align}
 \frac{g^2}{(g^2\mu)^4}\cdot\varepsilon &\simeq
  \frac{1}{2}N_c(N_c^2-1)\,\sigma\biggl[\sigma-\frac{1}{2}
  \frac{(g^2\mu\tau)^2}{(g^2\mu a)^2}\biggr]
\label{eq:result} \,,\\
 &\simeq 12\Biggl\{\frac{1}{4\pi}\ln\biggl[\frac{c^2(g^2\mu L)^2}
  {(g^2\mu a)^2+\pi(g^2\mu\tau)^2}\biggr]\Biggr\}^2 \,,
\label{eq:resum}
\end{align}
and
\begin{equation}
 \varepsilon = \frac{3(g^2\mu)^4}{\pi^2 g^2}\Bigl\{\ln
  \bigl(\LQCD^{-1}/\tau\bigr)\Bigr\}^2 \,.
\label{eq:compact}
\end{equation}
We remark that the first term of above Eq.~(\ref{eq:result}) has the
same overall factor as given in Eqs.~(14) and (16) in
Ref.~\cite{Lappi:2006hq}.  Our calculations~\cite{future} make use of
the technique developed in Ref.~\cite{Blaizot:2004wv} (the necessary
correlation function is obtained from $\mathcal{N}^{(b)}$ derived as
Eq.~(75) in Ref.~\cite{Blaizot:2004wv}), while Eqs.~(16) in
Ref.~\cite{Lappi:2006hq} seems to be based on Eq.~(2.23) in
Ref.~\cite{JalilianMarian:1996xn}.  These are independent calculations
with different cut-off prescriptions, but of course, the coefficients
in front of the logarithmic singularity should coincide with each
other.

  In a quantitative sense, in fact, the analyses on the Glasma
instability should be affected by the missing factor $16$ through
Eqs.~(22) and (23) in Ref.~\cite{Fukushima:2007ja}.  Our discussion on
instability is not beyond the qualitative level, however, and the
essential idea for the possible instability mechanism still works.
Also, in the first part of Ref.~\cite{Fukushima:2007ja}, no
modification is required in the essential ideas;  the energy density
at $\tau=0$ has a logarithmic divergence and the expansion in terms
of $\tau$ is singular around $\tau=0$.  The resummed form in the
logarithmic ansatz is a natural choice to remove the singularity at
finite $\tau$.\\


  One question arises immediately, however.  The initial energy
density has been evaluated in the \textit{same} model and the
\textit{same} cut-off prescription in different two methods.  If both
methods equally work well, two results must be identical apart from a
small discrepancy originating from the lattice and continuum
formulations.  Why can one differ from the other?

  This problem is so interesting on its own that it may deserve
another publication~\cite{future}, but we shall briefly see where the
difference stems from.  In short, the important point is that the
McLerran-Venugopalan (MV) model implemented in the numerical
calculation is not faithful to the analytical formulation.  The
subtlety originates from how to define the ill-defined expression of
Eq.~(4) in Ref.~\cite{Fukushima:2007ja}.  We should introduce some
regularization to write it in a form of
\begin{equation}
 V_\epsilon^\dagger(\xt) = \Pcal\exp\biggl[
  -ig\int\!dz^- \frac{1}{\del^2} \rho_\epsilon(\xt,z^-) \biggr] \,.
\label{eq:wilson}
\end{equation}
Here we defined the regularized color source as
\begin{equation}
 \lim_{\epsilon\to0}\rho_\epsilon(\xt,x^-)
  = \rho(\xt)\,\delta(x^-) \,.
\end{equation}

  Another delta function in the longitudinal direction appears in the
correlation function in terms of the sources,
\begin{equation}
 \begin{split}
 &\bigl\langle\rho_a(\xt,x^-)\,\rho_b(\yt,y^-)\bigr
  \rangle_\zeta \\
 &\qquad\qquad = g^2\mu^2(x^-)\,\delta_{ab}\,\delta(\xt\!-\!\yt)
   \delta_\zeta(x^-\!-\!y^-) \,,
 \end{split}
\end{equation}
where we replaced the delta function by the regularized one in the
longitudinal direction such that,
\begin{equation}
 \lim_{\zeta\to0}\delta_\zeta(x^-) = \delta(x^-) \,.
\end{equation}

  The question we are addressing here is whether we are allowed to
adopt the following simplification;
\begin{equation}
 V^\dagger(\xt) \overset{?}{\to} \:
 V^\dagger(\xt) = \exp\biggl[ -ig\frac{1}{\del^2}
  \rho(\xt) \biggr] \,,
\label{eq:approx}
\end{equation}
to prevent the delta function from appearing at all.  The numerical
calculations commonly make use of Eq.~(\ref{eq:approx}) for practical
reasons with hope that the final answer would not depend on this
replacement (see, e.g.\ Eq.~(36) in Ref.~\cite{Krasnitz:1998ns},
Eq.~(4) in Ref.~\cite{Lappi:2003bi}, and so on).


  Using the notations we introduced above, we can reiterate the
question more rigorously.  That is, we shall check,
\begin{equation}
 \lim_{\zeta\to0}\lim_{\epsilon\to0}
  \bigl\langle \mathcal{O}\bigl[ V_\epsilon\bigr]\bigr\rangle_\zeta
 \overset{?}{=}
 \lim_{\epsilon\to0}\lim_{\zeta\to0}
  \bigl\langle \mathcal{O}\bigl[ V_\epsilon\bigr]\bigr\rangle_\zeta \,.
\label{eq:limits}
\end{equation}
The left-hand side corresponds to the numerical implementation and the
right-hand side to the analytical formulation of the MV model.

  We treat more general cases in a separate literature~\cite{future}
and will limit the current discussion only to the tadpole calculation,
namely, $\mathcal{O}[V]=V^\dagger$.  This simplest choice is, as we
will see, enough to exemplify two limits in Eq.~(\ref{eq:limits}) are
noncommutative.

  We already know the analytical answer for the right-hand side of
Eq.~(\ref{eq:limits}).  That is given
by~\cite{Blaizot:2004wv,Fukushima:2007dy}
\begin{equation}
 \lim_{\epsilon\to0}\lim_{\zeta\to0}\bigl\langle V_\epsilon^\dagger
  \bigr\rangle_\zeta = \exp\biggl[-g^4\mu^2\frac{N_c^2-1}{4N_c}\,
  \eta\biggr] \,,
\label{eq:lhs}
\end{equation}
where $\mu^2=\int dx^-\mu^2(x^-)$ and
\begin{equation}
 \eta = \frac{a^2}{4L^2}\sum_{n_i=1-L/2}^{L/2}\frac{1}
  {\bigl[2\!-\!\cos(2\pi n_1/L)\!-\!\cos(2\pi n_2 /L)\bigr]^2} \,.
\end{equation}

  To evaluate the left-hand side, we have to perform the following
Gaussian integral,
\begin{align}
  &\lim_{\zeta\to0}\lim_{\epsilon\to0}
  \bigl\langle V_\epsilon^\dagger\bigr\rangle_\zeta
  = \int[d\rho] \exp\biggl[-ig\frac{1}{\del^2}\rho_a(\xt)\,t^a
  \biggr] \notag\\
  &\qquad\qquad\times\exp\biggl[-\int d^2\xt \frac{\rho_a(\xt)\rho_a(\xt)}
  {2g^2\mu^2}\biggr] \,,
\end{align}
where $t^a$'s are the SU($N_c$) algebra.  It is hard to do this
integration for arbitrary SU($N_c$) group, while the SU(2) case is
immediately doable.

  In the case of SU(2) (i.e.\ $N_c=2$), the above Gaussian integral
leads to
\begin{equation}
 \lim_{\zeta\to0}\lim_{\epsilon\to0}
  \bigl\langle V_\epsilon^\dagger\bigr\rangle_\zeta
  =\biggl(1-g^4\mu^2\frac{1}{4}\eta\biggr)
  \exp\biggl[-g^4\mu^2\frac{1}{8}\eta\biggr] \,,
\label{eq:rhs}
\end{equation}
\vspace{3mm}

\noindent
which is obviously different from the right-hand side with $N_c=2$
substituted, that is,
\begin{equation}
 \lim_{\epsilon\to0}\lim_{\zeta\to0}
  \bigl\langle V_\epsilon^\dagger\bigr\rangle_\zeta
  =\exp\biggl[-g^4\mu^2\frac{3}{8}\eta\biggr] \,,
\label{eq:lhs2}
\end{equation}
from Eq.~(\ref{eq:lhs}).

  Interestingly, though Eqs.~(\ref{eq:rhs}) and (\ref{eq:lhs2}) look
quite different, the curvature near $\eta\simeq0$ is the same.  As we
report in Ref.~\cite{future}, the discrepancy between the left-hand
and right-hand sides in Eq.~(\ref{eq:limits}) becomes significant when
the singlet component survives unsuppressed by the system size.  In
fact, the expectation value of the gauge fields,
$\langle V_\epsilon \partial_i V_\epsilon^\dagger\cdot
V_\epsilon \partial_i V_\epsilon^\dagger\rangle_\zeta$, leads to
different energy densities by a factor.

\end{document}